\documentclass[prl,twocolumn,nofootinbib]{revtex4}
\usepackage{graphicx}
\usepackage{color}
\usepackage{epsfig}
\usepackage[caption=false]{subfig}
\usepackage{grffile}
\usepackage{multirow}
\begin{document}

\title{Possible Universal Relation Between Short time $\beta$-relaxation and 
Long time $\alpha$-relaxation  in Glass-forming Liquids}
\author{Rajsekhar Das\footnote{These authors contributed equally.}}
\email{rajsekhard@tifrh.res.in}
\author{Indrajit Tah$^{*}$}
\email{indrajittah@tifrh.res.in}
\author{Smarajit Karmakar}
\email{smarajit@tifrh.res.in}
\affiliation{
$^1$ Tata Institute of Fundamental Research, 
36/P, Gopanpally Village, Serilingampally Mandal,Ranga Reddy District, Hyderabad, 500107, India}

\begin{abstract}
Relaxation processes in supercooled liquids are known to exhibit interesting 
as well as complex behavior. One of the hallmarks of this relaxation 
process observed in the measured auto correlation function is occurrence of 
multiple steps of relaxation. The shorter time relaxation is known as the 
$\beta$-relaxation which is believed to be due to the motion of particles in 
the cage formed by their neighbors. One the other hand longer time relaxation, 
the $\alpha$-relaxation is believed to be the main relaxation
process in the liquids. The timescales of these two relaxations processes
dramatically separate out with supercooling. In spite of decades 
of researches, it is still not clearly known how these relaxation processes are 
related to each other. In this work we show that, there is a possible universal 
relation between short time $\beta$-relaxation and the long time 
$\alpha$-relaxation. This relation is found to be quite robust across many 
different model systems. Finally we show that length scale obtained 
from the finite size scaling analysis of $\beta$ timescale is same as that of 
length scale associated with the dynamic heterogeneity in both two and three 
dimensions.
\end{abstract}

\maketitle
Dynamics of supercooled liquids are very complex in nature. The decay of two
point density-density auto correlation shows two steps relaxations as the 
liquid is supercooled. In spite of decades of research the complex dynamical 
behaviours associated with putative glass transition, is still poorly 
understood \cite{11BB,BB2012, arcmp, KDSROPP16}. The density auto correlation 
function decays to a plateau at 
shorter time and then at much longer time it finally decays from the 
plateau to zero in a stretched exponential manner. The short time relaxation 
in the plateau region is known as $\beta$-relaxation whereas the longer 
time relaxation is called the $\alpha$-relaxation \cite{11BB,BB2012,JG,MEMK2009, SK2016}. Although a lot of efforts 
have been made to understand the nature of $\alpha$ relaxation and 
its microscopic origin \cite{11BB,BB2012, arcmp, KDSROPP16}, far less 
researches are done to understand the same for shorter time $\beta$-relaxation 
and its possible connection with the $\alpha$-relaxation \cite{steinAdersen, 
KDS2016,NGAI98, HWHYK2014, NGAI2013}. 

At short times, it is believed that the particles get trapped in transient 
cage formed by their neighboring particles and they undergo a kind of 
rattling motion in those cages. Eventually they hop out of the cage and 
probably after successive such cage breaking processes, the liquid finally
relaxes. It is also not clearly known whether the rattling in a cage and 
subsequent breaking of the cage is the $\beta$-relaxation. If one assumes 
such an event to be a $\beta$-relaxation and multiple such events leads to 
$\alpha$-relaxation, then one can expect that short time and long time
relaxation processes will be intimately related to each other. Such a 
scenario is indeed suggested in some studies \cite{NGAI98, NGAI2013, 
KDS2016}. 

In \cite{NGAI98}, author have proposed a correlation between 
$\beta$-relaxation time $\tau_\beta(T_g)$ calculated at the experimental glass 
transition temperature ($T_g$),  and the 
Kohlrausch-Williams-Watts (KWW) exponent $(1-n)$ of $\alpha$-relaxation 
at $T_g$. $T_g$ is defined experimentally at the temperature where 
the relaxation time of the liquid becomes $100s$. The KWW exponent 
is the stretching exponent of the decay profile of the two-point 
density-density auto correlation function or the self intermediate 
scattering function as 
\begin{equation}
Q(t) = \exp{\left[ -\left(\frac{t}{\tau_{\alpha}}\right)^{1-n}\right]},
 \quad 0\le n \le 1,
\end{equation}
where $Q(t) = \left<\frac{1}{N}\sum_{i=1}^{N}w(|\vec{r}_t(t) - \vec{r}_i(0)|)\right>$. The window 
function $w(x) = 1.0$ if $x \leq 0.3$ and $0$ otherwise
(see SI for further details). 
%This window function is used to remove possible decorrelation that 
%might happen due to vibrational motion inside the cage formed around 
%a particle in a dense system by their neighbors.  
$\langle \ldots\rangle$ 
represents ensemble average. The $\alpha$ relaxation time is 
defined as $Q(\tau_\alpha) = 1/e$. 
At high temperature, the relaxation in liquid is exponential, 
that is $n =0$, thus one can expect that the nonzero value of 
$n$ will be related to many body or cooperative nature of the 
$\alpha$-relaxation process. 

The Coupling Model (CM) proposed in \cite{NGAI1979,NGAI1994,Tsang1997}, 
suggests a relation between primitive relaxation time, $\tau_0$ with that 
of the the $\alpha$-relaxation time as
\begin{equation}
\tau_\alpha = \left[\tau_0\tau_c^{-n} \right]^{1/(1-n)},
\end{equation}
where $\tau_c$ \cite{ZACFRB1995,CAA1993} is the microscopic time scale. 
$\tau_0$, the primitive relaxation time is argued to be close to 
$\beta$-relaxation time as both of them are assumed to be the 
precursors of the long time $\alpha$-relaxation process. This 
relationship has been tested for many glass forming liquids near 
the experimental glass transition temperature and found to agree 
with the above relation to varying degree. Fujimori and Ouni's 
correlation index $c$ \cite{Fujimori1995}, defined as 
$c \equiv 1- \frac{T_{g\beta}}{T_{g\alpha}}$,  and the coupling 
parameter $n$ of CM was also shown subsequently to be linearly 
proportional to each other for many experimental glass-forming 
liquids. This clearly suggests that although there are proposals
and reports of possible inter-relation between $\tau_\alpha$
and $\tau_\beta$, a general consensus is still missing.

In a recent study \cite{KDS2016}, it has been shown that the 
system size dependence 
of $\tau_{\beta}$ in three dimensional glass-forming liquids is 
controlled by the dynamic heterogeneity length ($\xi_d$) that are 
obtained from the finite size scaling\cite{KDS2009,Privman1990} of 
peak height of the four-point dynamic susceptibility ($\chi_4(t)$, 
see SI for definition), $\chi_4^P(T)$\cite{arcmp,KDS2009}. 
The peak of $\chi_4(t)$ 
appears at $\alpha$-relaxation time scale, suggesting a very strong 
inter-relation between these two relaxation processes. Thus a 
possible universal relation between $\alpha$ and $\beta$ time 
scale and its origin can be connected to the growth of different 
length-scales in the system. The main objective of the present 
work is to revisit this possible relationship between $\tau_\beta$ 
and $\tau_{\alpha}$ and try to explore existence of an universal 
relationship between these two timescales using more microscopic 
quantities like dynamic heterogeneity length scale ($\xi_d$) and 
static length scale ($\xi_s$) that grow with supercooling 
\cite{arcmp,KDSROPP16,KDS2014,BBMR2006}.

In this article, we propose a new universal relation between $\beta$-relaxation 
and $\alpha$-relaxation in model glass-forming liquids and try to 
rationalize the results within the framework of the well-known Random 
First Order Transition (RFOT)\cite{KTW1989,LW2007,BB2012} theory of glass 
transition. Rest of the paper is arranged as follows. First we briefly 
discuss some of the details of the simulation methods and the models
and then we define some relevant correlation functions that are used 
to calculate  different relaxation times and length-scales. A set of 
new quantities are defined to analyze the data particularly for two 
dimensional systems. In two dimensions, there will be contribution 
from long-wave length phonon mode and appropriate corrections need 
to be made to disentangle the effect due to glass transition and long 
wavelength density fluctuations on the measured quantities 
\cite{szamelNatComm, KABetaPaper, d1234glass, weeksPNAS, KeimPNAS, 
gilles2d3d,SYKK2016}.  
We then discuss about the relation between $\tau_{\alpha}$ and 
$\tau_{\beta}$ for both two and three dimensional systems. Effects 
of finite size on these results are then discussed. Finally we 
rationalize our observation within the framework of RFOT theory 
and propose a new universal relationship between $\tau_{\alpha}$ 
and $\tau_{\beta}$.

We have studied different model systems in two and three dimensions with 
somewhat different inter particle potentials to make sure that the 
results obtained are generic and applicable for wide variety of 
systems. First model is the well known Kob-Andersen binary model where 
particles interact via Lennard-Jones potential. We refer the model as 
{\bf 3dKA}\cite{KA1995}. The second model studied is also a binary 
mixture of particles but with pure repulsive inter particle interactions 
and this is referred here as {\bf 3dR10}\cite{KLPZ2010}. Other models 
 are {\bf 3dIPL\cite{PSD2010}, 3dHP\cite{OLLN2002,blockprl}, 
3dBMLJ\_82}\cite{cos2007}. We have done very large system size simulations
in three dimensions to remove finite size effects (see SI for further 
details). We have done simulations for system sizes in the range 
$N\in {[1000, 108000]}$. The data reported in the article is for 
$N = 108000$ only.   
In two dimensions, we study same Kob-Andersen model and refer it as
{\bf 2dKA}. A slightly modified version of 2dKA model is also studied and 
will be referred as {\bf 2dmKA} model. The two dimensional version 
of 3dR10 model will be referred as {\bf 2dR10}. A model of
polydisperse mixture of particles with truncated Lennard-Jones 
inter particles interaction potential (also known as WCA potential) 
in two dimensions is also studied. We refer that model as 
{\bf 2dPoly}. The system size in two dimensions ranges from $N\in [100,10000]$.    
All the details regarding these 
different models and the simulations details are given in the SI.
\begin{figure}
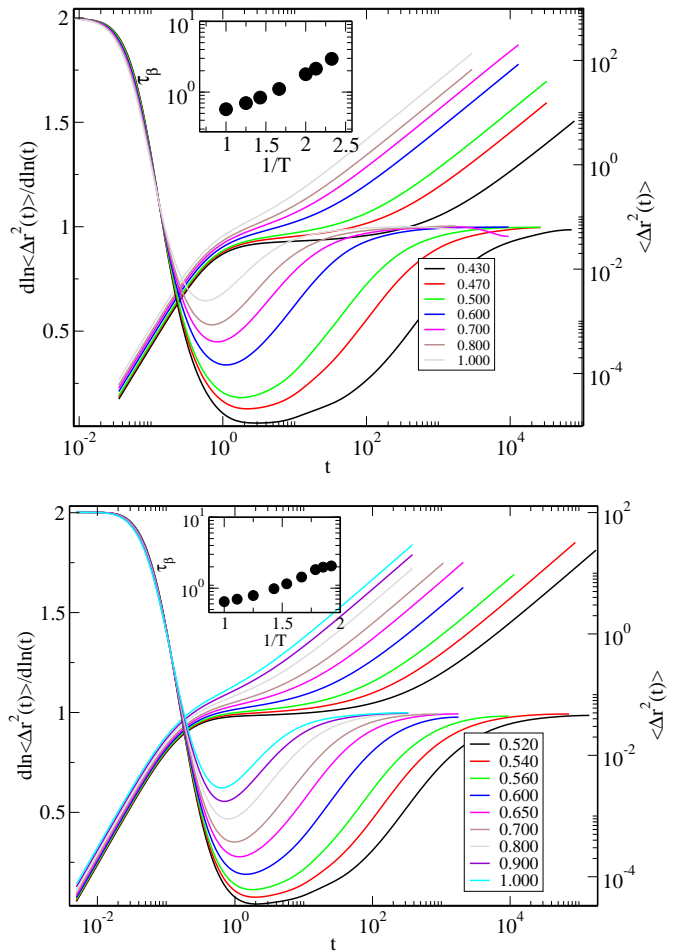

\begin{center}
\vskip +0.2cm
\includegraphics[width=0.48\textwidth]{msdlogderivmsd_3dKA_with_inset.eps}\\
\vskip +0.3cm
\includegraphics[width= 0.49\textwidth]{msdlogderivmsd_3dR10.eps} 
\caption{Top panel: Mean square displacement(MSD) and its log-derivative 
are plotted against time for 3dKA model. In the inset we have shown the 
temperature dependence of $\tau_{\beta}$ for the same model. The 
temperature dependence is close to Arrhenius in agreement with previous 
experimental results. Bottom panel: Similar plot for 3dR10 model. 
In the inset we show the temperature dependence of $\tau_\beta$ for 
3dR10 model. For this model also the temperature dependence of 
$\tau_\beta$ follows Arrhenius behavior.}
\vskip -0.4cm
\label{MSD_LOGMSD_3d}
\end{center}
\end{figure}

\vskip +0.15cm
\noindent{\bf Calculation of $\beta$ time scale:}
We follow the method given in \cite{steinAdersen,KDS2016} to calculate 
$\beta$-relaxation timescale, $\tau_\beta$ from the mean squared 
displacement (MSD) which is defined below as  
\begin{equation}
\langle |\Delta r(t)|^2 \rangle = \left\langle \frac{1}{N}\sum_{i=1}^N 
|\vec{r}_i(t) - \vec{r}_i(0)|^2\right\rangle.
\end{equation}
MSD shows a point of inflection at an intermediate time, thus if we 
plot the log-derivative of MSD with time, 
$d\log{\langle |\Delta r(t)|^2 \rangle}/d\log{t}$, it will 
show a dip at that inflection point. $\tau_{\beta}$ is defined as the 
time where the point of inflection appears. This procedure 
is shown for 3dKA (top panel) and 3dR10 (bottom panel) models in 
Fig.\ref{MSD_LOGMSD_3d}. One can clearly see the minimum
in $d\log{\langle |\Delta r(t)|^2 \rangle}/d\log{t}$ vs $t$ plots 
and also the minimum shifts to higher and higher values of $t$ with 
decreasing temperature. 

\begin{figure}[h]
\begin{center}
\hskip -0.742cm
\includegraphics[width=0.48\textwidth]{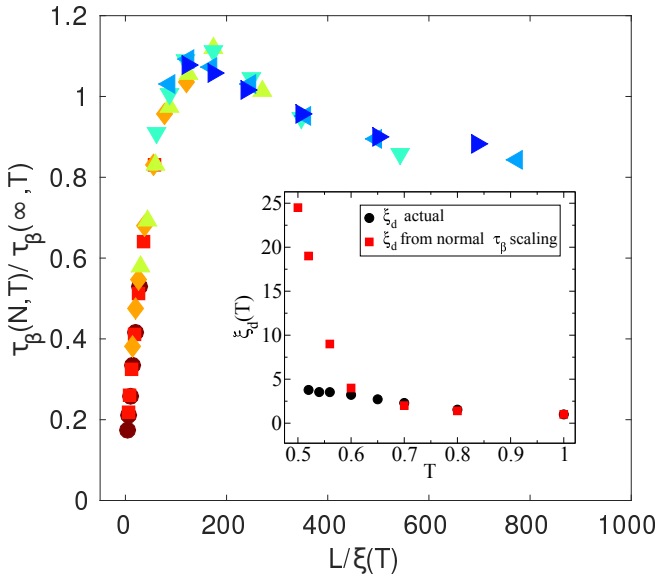}
%\vskip -0.40cm
\caption{Data collapse of finite size dependence of $\tau_\beta$ for 
{\bf 2dR10} mode. Here $\tau_\beta$ is obtained from normal MSD (see text
for details). Inset shows the variation of length scale with temperature. 
The scale obtained here are very different than the dynamic heterogeneity 
length scale of this model system.} 
\vskip -0.5cm
\label{FSS_NORMAL_MSD_2dR10}
\end{center}
\end{figure}
In \cite{KDS2016}, it was shown that the system size dependence of 
$\tau_\beta$ is controlled by the dynamic heterogeneity length scale 
in three dimensions. To check the validity of the same results in 
two dimensional models, we have done the finite size scaling (FSS) 
analysis of $\tau_\beta$ for 2dR10 model. The results are shown in
Fig.\ref{FSS_NORMAL_MSD_2dR10}. $\tau_\beta$ has fairly large system
size dependence (shown in SI) and the dependence becomes very strong 
at lower temperatures. Although the data collapse observed is quite 
good, the obtained length scales as shown in the inset, is found  to be
very different from the dynamic heterogeneity length scale of this model 
obtained via different methods. 

\begin{figure}
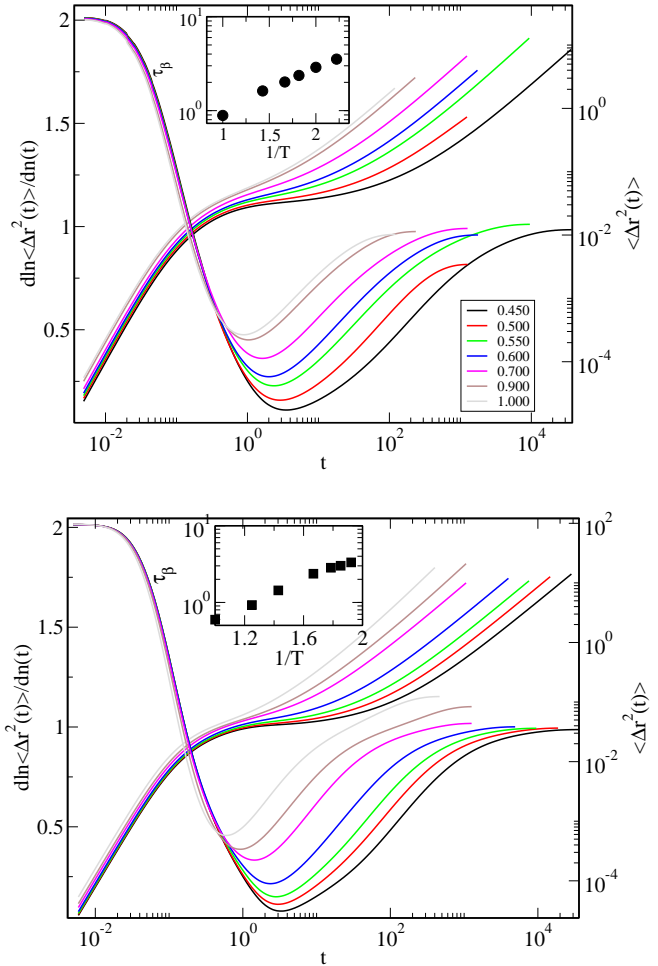

\begin{center}
\vskip +0.5cm
\includegraphics[width=0.470\textwidth]{msdlogderivmsd_ds_2dmKA_with_inset.eps}
\vskip +0.5cm
\includegraphics[width=0.480\textwidth]{msdlogderivmsd_ds_2dR10_withinset.eps}\\ 
\hskip -0.8cm
\caption{ Top panel: Cage-relative mean square displacement(crMSD) and 
its log-derivative is plotted against time for 2dMKA system. Inset shows 
the Arrhenius temperature dependence of $\tau_\beta$. Bottom panel: Similar 
plot for the 2dR10 model. Similar Arrhenius temperature dependence of 
$\tau_\beta$ is shown for this model in the inset.} 
\label{crMSD}
\end{center}
\end{figure} 
The variation of the length scale in the studied temperature range
is very large suggesting a possible contribution coming from long wavelength 
density fluctuations which are prevalent in two dimensional systems due to 
Marmin-Wagner theorem\cite{MW1966,M1968,BSHCGP2017,KeimPNAS}. Thus disentangling 
contributions coming from these long wavelength fluctuations and the 
glass transition is very important to understand glass transition in two 
dimensions \cite{SYKK2016,SFGP2009,BSHCGP2017}. 
To overcome such a problem in two dimensional system we have calculated 
cage-relative MSD (crMSD) following \cite{SYKK2016,SFGP2009,BSHCGP2017}. 
The crMSD is defined as follows. First we define the cage related 
displacement of particle $i$ as
\begin{equation}
{\Delta}r_{i,CR}(t) = {\Delta}r_i(t) -\frac{1}{N_{nn}}{\sum_{j\in n.n} {\Delta}r_j(t)},
\end{equation}
where $N_{nn}$ are the number of nearest neighbor of $i^{th}$ particle, and 
${\Delta}r_j(t) = r_j(t) - r_j(0).$ Particles are defined as the neighbors of 
$i^{th}$ particle if they satisfy $r_{ij}(t) = |r_j(t) - r_j(0)|< A\sigma_{\alpha\beta}$. 
$A$ is the cutoff used to define the neighbors. It is usually taken as the 
distance where first minimum of the radial distribution function $g(r)$ 
appears. The  crMSD is then defined as,
\begin{equation}
\left\langle |{\Delta}r_{CR}(t)|^2 \right\rangle = \left\langle \frac{1}{N} \sum_{i = 1}^{N} |{\Delta}r_{i,CR}(t)|^2 \right\rangle
\end{equation} 
\begin{figure}
\begin{center}
\includegraphics[width=0.5\textwidth]{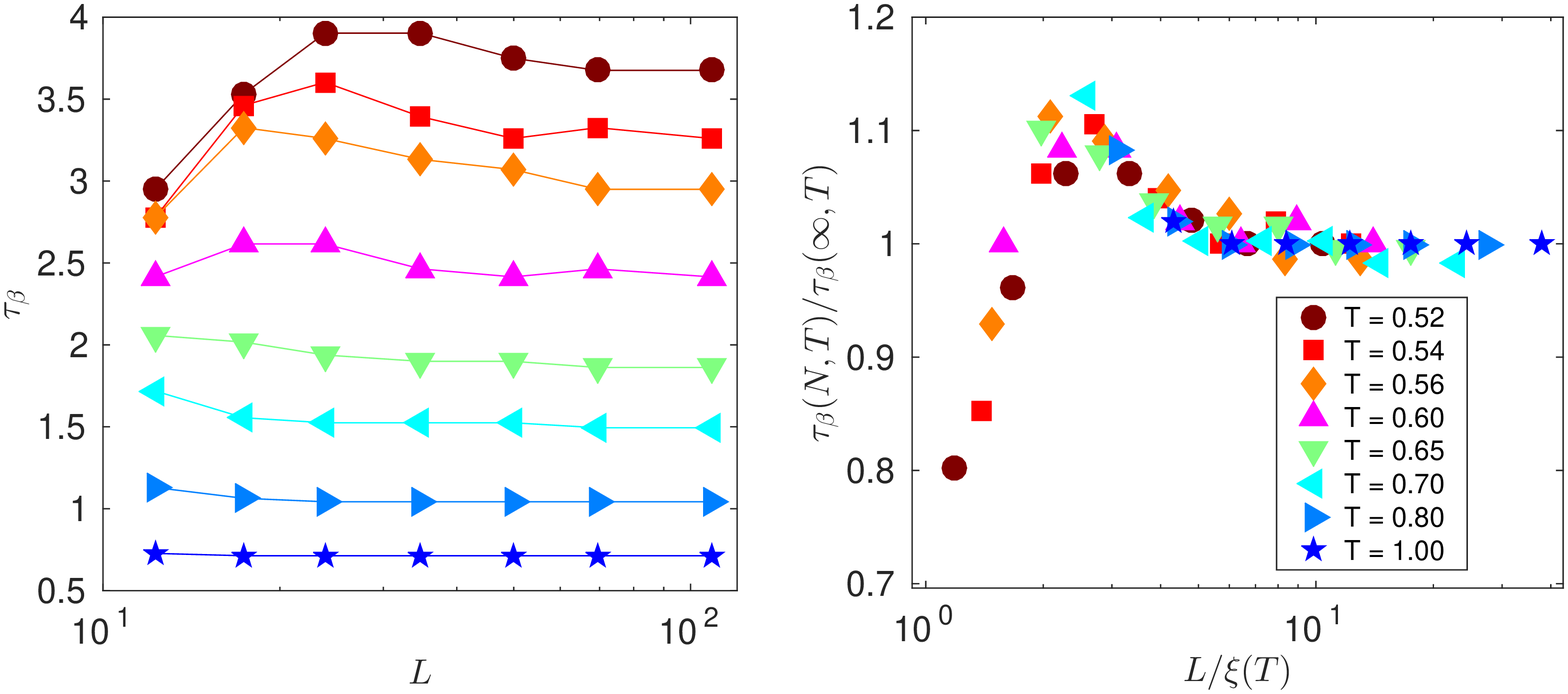}
\caption{ Left panel: System size dependence of $\tau_\beta$ obtained from the
log-derivative of cage-relative mean square displacement(crMSD) with dynamic 
heterogeneity length scale as the cutoff (see text for details) for the
2dR10 model. Right panel: Data collapse to obtain the length scale which is 
found to be same as the dynamic heterogeneity length scale.} 
\vskip -0.80cm
\label{tauBetaScaling2dR10}
\end{center}
\end{figure}
As we are measuring relative displacements, it will not be affected by 
the long wavelength phonon modes and thus one should be able to extract 
the relevant information from MSD related only to glass transition in 
two dimensions. 

It turns out that the results very much depend on how the neighboring 
particles are chosen, for example, if one chooses a cutoff at the
second minimum in $g(r)$, then one finds a week system size dependence 
of $\tau_\beta$ compare to almost no system size dependence if first 
minimum of $g(r)$ is chosen. This is somewhat puzzling and seems to 
constraint the usefulness of the crMSD. One can rationalize 
these results from the understanding that if one defines a cage relative motion 
using particles which are its immediate neighbors then one is basically 
removing even local cooperative motions in the systems. This local cooperative 
motion has nothing to do with the long wavelength phonon mode. 

To keep the cooperative motions undisturbed over 
the dynamic heterogeneity length scale, we choose the cutoff length for 
defining the neighbors to be same as the dynamic heterogeneity length. 
This procedure
gives us an estimate of $\tau_\beta$ which is not affected by the long 
wavelength density fluctuations at the same time any possible 
contributions coming from cooperative motions will not be washed away. 
In our subsequent analysis we have followed this method to calculate 
$\tau_\beta$. In the top panels of Fig.\ref{crMSD}, we have shown 
$d\log{\langle |\Delta r_{CR}(t)|^2 \rangle}/d\log{t}$ vs $t$ plots for 2dmKA 
and 2dR10 models. In the inset we show the Arrhenius temperature dependence of 
$\tau_\beta$ for these model systems. 
\begin{figure}
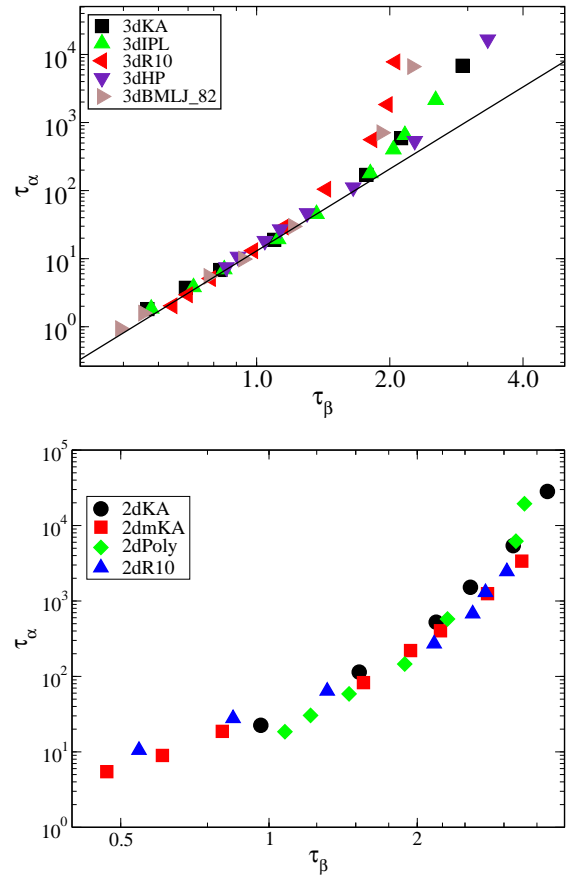

\includegraphics[width=0.85\columnwidth]{3d108K_my.eps}
\vskip +0.2cm
\includegraphics[width=0.85\columnwidth]{log_tau_alpha_log_tau_beta_ds_2dALL.eps}
\caption{Log-log plot of $\tau_\alpha$ vs $\tau_\beta$ for all the models systems
in three dimensions (top panel) and two dimensions (bottom panel). Power law 
relation seems not to be valid for all the temperature and data for different
models deviate from the master curve at lower temperature in 3d dimensional 
models. Some quasi-universal relation is observed in two dimensions.}
\label{CMcheck}
\vskip -0.5cm
\end{figure}

In the left panel of Fig.\ref{tauBetaScaling2dR10}, we show the system 
size dependence of $\tau_\beta$ for 2dR10 model and in the right panel we show 
the finite size scaling of the same data using dynamic heterogeneity length 
scale, $\xi_d$ taken from Ref.\cite{ISSCS17}. The data collapse is indeed  
reasonable. Thus it can be concluded that finite size scaling of 
$\tau_\beta$ is governed by the dynamic heterogeneity length scale in
two dimensions also. This is very similar to the observation reported for 
three dimensional model \cite{KDS2016}. This suggests that glass transitions 
in two and three dimensions share features which are very similar to each 
other if the effect of long wavelength phonon mode in two dimensions can 
be disentangled from the measured quantities. 
Similar observations were made in recent works 
\cite{weeksPNAS, KeimPNAS, gilles2d3d,SYKK2016,MEMK2009,BSHCGP2017}.  

After calculating $\tau_\beta$ and $\tau_\alpha$ for all the models in 
two and three dimensions reliably, we would like to focus our attention 
on the possible universal relation between these two timescales. As 
discussed earlier, in Ref.\cite{NGAI98}, a power law relationship has 
been proposed between $\tau_\alpha$ and $\tau_\beta$ as
$\tau_\alpha \sim \tau_\beta^{1/(1-n)}$. As KWW stretching exponent 
$1-n$ decreases from $1$ with supercooling in a manner which is 
similar for different model systems, one expects to be able to obtain 
a master curve by plotting $\tau_\alpha$ as a function of $\tau_\beta$ 
for all the temperatures with appropriate choice of the pre-factor in 
the power law relation. In Fig.\ref{CMcheck}, we have tested the same 
proposal by plotting $\tau_\alpha$ as a function of $\tau_\beta$ in 
log-log plot for both two and three dimensional models. 
\begin{figure}
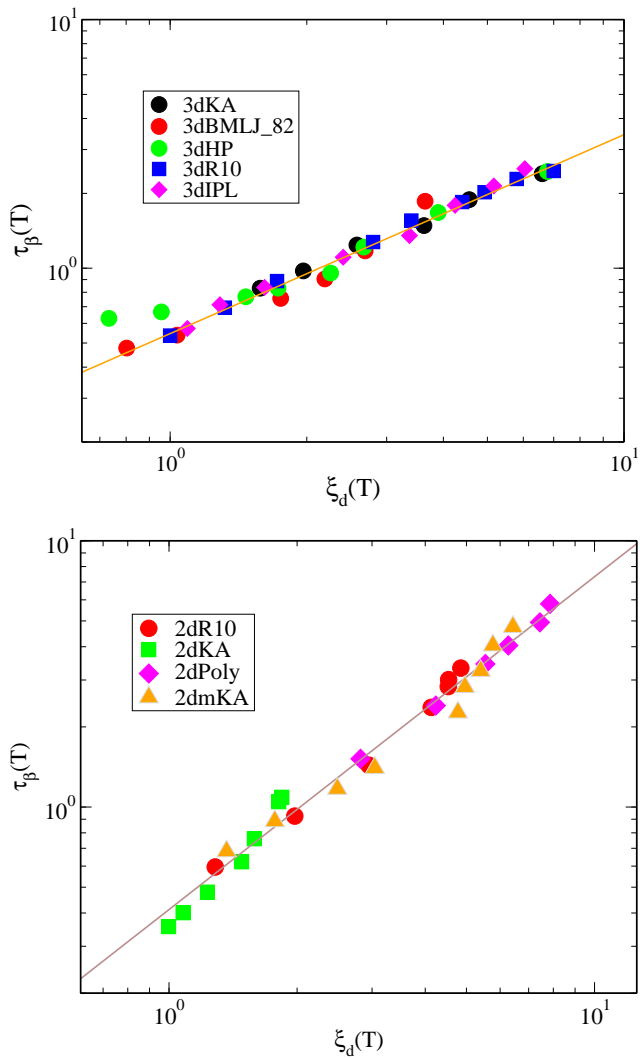

\begin{center}
\includegraphics[width=0.97\columnwidth]{tau_beta_xid_3dall_scaled.eps}
\vskip +0.2cm
\includegraphics[width=0.97\columnwidth]{taubetaxid_2d_1.eps}
\caption{Dynamic scaling relation between $\tau_\beta$ and $\xi_d$, 
$\tau_\beta \sim \xi_d^z$. Top panel shows data for all three dimensional 
model systems with $z \simeq 0.8$ and bottom panel shows data for two 
dimensional models with $z \simeq 1.25$.}
\label{taubetaxid}
\end{center}
\end{figure}

\begin{figure}[b]
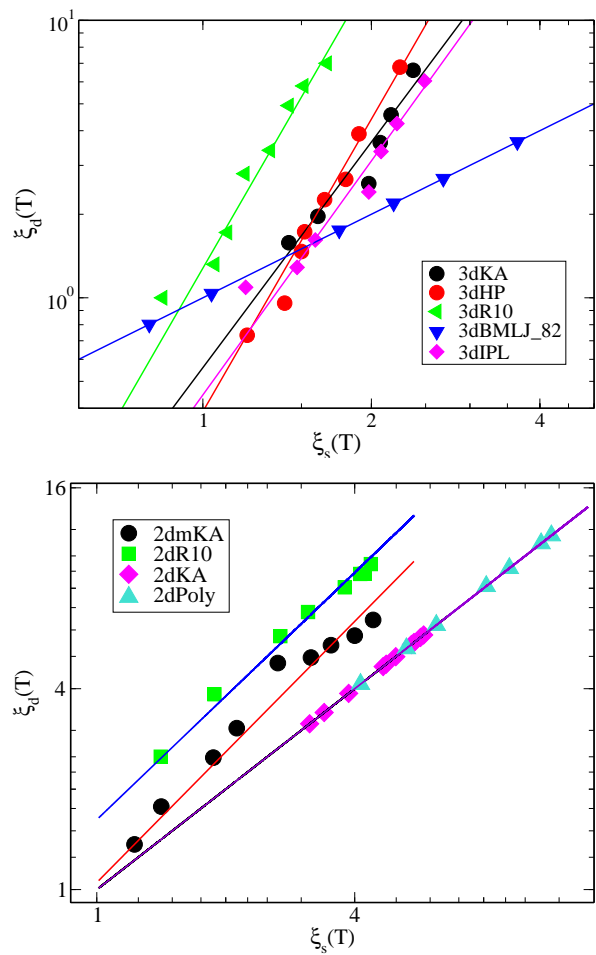

\begin{center}
\includegraphics[width=0.9\columnwidth]{static_dynamic_3dall.eps}
\vskip +0.2cm
\includegraphics[width=0.9\columnwidth]{xidxis_2d.eps}
\caption{Top panel: Relation between $\xi_d$ vs $\xi_s$, as $\xi_d \sim \xi_s^X$, 
with $X = 2.7$(3dKA), $3.5$(3dHP), $3.5$(3dR10), $2.8$(3dIPL), $1.0$(BMLJ82) 
for different three dimensional models. Bottom panel: similar analysis for
different two dimensional models with $X = 1.30$(2dmKA), $1.23$(2dR10), 
$1.00$ (2dKA and 2dPoly).}
\label{xidxis}
\end{center}
\end{figure}
\begin{figure*}
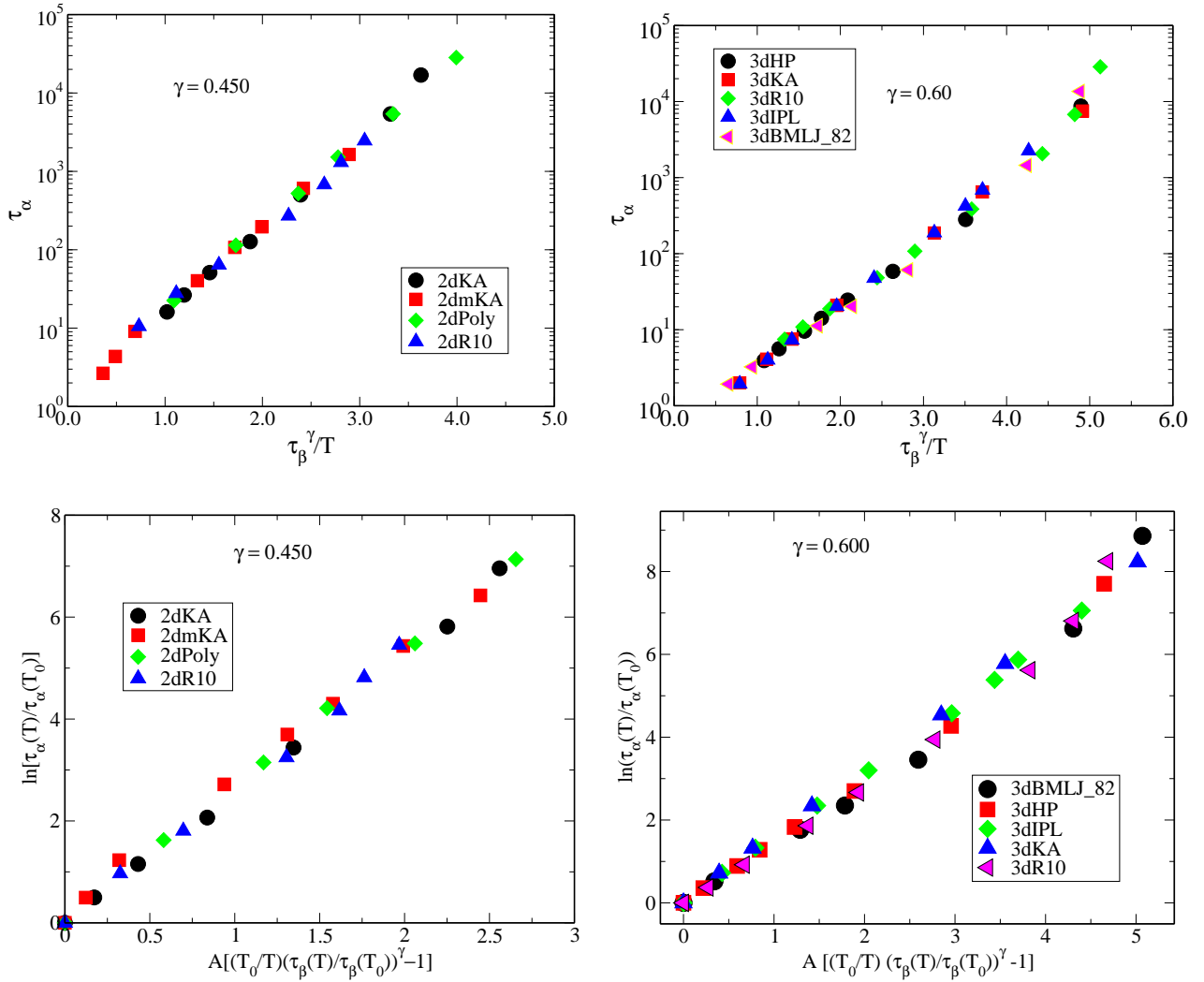

\begin{center}
\includegraphics[width=0.92\columnwidth]{logtaualph_vs_taubetaPowgammbyT_2dall_ds.eps}
\hskip +0.5cm
\includegraphics[width=0.950\columnwidth]{logtau_taubPowgammabyT3D_updated.eps}
\vskip +0.5cm
\includegraphics[width=0.92\columnwidth]{argument_1_all_2d_0.450_ds.eps}\hskip +0.5cm
\includegraphics[width=0.92\columnwidth]{newaegument_3dallupdated_gamm0.6.eps}
\caption{Top left panel: $\tau_\alpha$ is plotted against 
$\tau_\beta^{\gamma}/T$ for all the studied two dimensional model 
systems. Top right panel : Similar plot for all the studied three dimensional 
model systems. Both in two and three dimensions data follow a master curve 
which suggests the observed dependence to be very robust. Bottom left panel: 
$\ln[\frac{\tau_\alpha(T)}{\tau_\alpha(T_0)}]$ is plotted against 
$A[(\frac{T_0}{T})(\frac{\tau_\beta(T)}{\tau_\beta(T_0)})^{\gamma} -1]$ for 
all the two dimensional model systems. All the data are clearly on a single 
master curve with $\gamma = 0.45$. Bottom right panel: Similar plot for 
all the model three dimensional model systems with $\gamma = 0.60$.}
\label{alphaBetaRelation2}
\end{center}
\end{figure*}

One can see that the power law relation is not very robust in the studied 
temperature range for three dimensional systems and deviates strongly especially
at lower temperatures. In two dimensions also one sees similar results but data
seem to fall on a quasi universal master curve. The observed data collapse 
although is not very satisfactory. In this work, we propose a simple but robust 
relation between $\alpha$-relaxation and $\beta$-relaxation time. The form of 
the proposed relationship between these two timescales can be rationalized within
the framework of Random First Order Transition (RFOT) theory 
\cite{KTW1989,LW2007,BB2012}. In RFOT long time 
structural relaxation time $\tau_\alpha$ is connected to static length scale, 
$\xi_s$ via,
\begin{equation}
\tau_\alpha(T) = \tau_{\infty}\exp{\left(\frac{{\mu}\xi_s^{\psi}}{K_BT}\right)}.  
\end{equation}
As shown in \cite{KDS2016} and in present work, finite size effects of 
$\tau_\beta$ can be understood using the dynamic heterogeneity length scale, 
$\xi_d$ in both dimensions, we expect from dynamic scaling arguments that 
$\tau_\beta$ will be related to $\xi_d$ as 
\begin{equation}
\tau_\beta \sim \xi_d^z,
\end{equation}
where exponent $z$ is found to be close to $0.8$ for all three dimensional model
systems and $z\simeq1.25$ for all the studied models in two dimensions as shown
in Fig.\ref{taubetaxid}.

If one assumes that static and dynamic length scales are related 
as $\xi_d \sim \xi_s^X$, then the following relation between $\tau_\alpha$ 
and $\tau_\beta$ can be obtained.
\begin{equation}
\tau_\alpha  = \tau_0 \exp{\left( \frac{\Omega \tau_\beta^\gamma}{K_BT}\right)},
\label{universalRel}
\end{equation}
where exponent $\gamma = \psi/zX$. Thus if $\tau_\alpha$ is plotted as a function
of $\tau_\beta^\gamma/T$, then one would expect to have a master curve if exponent 
$\gamma$ is universal for all model systems. In general, there is no reasons
to expect $\gamma$ to be universal as exponent $\psi$ and $X$ are somewhat 
different amongst different model systems (see SI for the values of 
$\psi$ for different models). As shown in Fig.\ref{xidxis}, exponent $X$ is 
indeed different for different models in both two and three dimensions. 
The exponent $X = 2.7$, $3.5$, $3.5$, $2.8$ and $1.0$ for 3dKA, 3dHP, 3dR10,
3dIPL and 3dBMLJ\_82 respectively. As the range of power law is somewhat 
restricted, reliable estimate of $X$ is not easy. The values of $X$ 
for different two dimensional models are $X = 1.30$(2dmKA), $1.23$(2dR10), 
$1.00$ (2dKA and 2dPoly). Note that,
there are model systems which show presence of prominent medium range 
crystalline order at lower temperature or higher density and for them 
it has been shown that static and dynamic length scales are same 
\cite{ISSCS17}. 3dBMLJ\_82 in three dimensions and 2dKA, 2dPoly in two 
dimensions are examples of such models. The 
data for $\xi_d$ and $\xi_s$ are taken from \cite{SRS16,ISSCS17}.  

We then test the universality of Eq.\ref{universalRel}, for different two 
and three dimensional model systems. In top panels of 
Fig.\ref{alphaBetaRelation2}, we have plotted $\tau_\alpha$ as a function 
of $\tau_\beta^\gamma/T$ with $\gamma = 0.60$ for three dimensions 
(top left panel) and $\gamma = 0.45$ for two dimensions (top right panel). 
In these plots we have adjusted the values of non-universal parameters 
like $\tau_0$ and $\Omega$ in Eq.\ref{universalRel} to collapse all
the data on a master curve. The quality of collapse clearly suggest that
indeed $\tau_\alpha$ is universally related to $\tau_\beta$ via 
Eq.\ref{universalRel} with a universal exponent $\gamma$.

Now one might suspect the reliability of the above relation as $\tau_0$ 
and $\Omega$ are varied freely, to eliminate dependence on $\tau_0$, we 
take a reference temperature $T_0$ (highest temperature studied for 
each model) and divide Eq.\ref{universalRel} from both sides to write,
\begin{equation}
\log\left[\frac{\tau_\alpha(T)}{\tau_\alpha(T_0)}\right] = 
\Omega\left[\left(\frac{T_0}{T}\right)\left(\frac{\tau_\beta(T)}{\tau_\beta(T_0)}\right)^{\gamma} -1\right]
\label{universalRelmod}
\end{equation}
Eq.\ref{universalRelmod}, keeps only one parameter free and that is 
$\Omega$. In bottom panels of Fig.\ref{alphaBetaRelation2}, we have plotted 
$\log\left[\frac{\tau_\alpha(T)}{\tau_\alpha(T_0)}\right]$ vs $\frac{T_0}{T}
\left(\frac{\tau_\beta(T)}{\tau_\beta(T_0)}\right)^{\gamma} -1$ for both two 
and three dimensions. The observed data collapse again reconfirms the proposed
universal relation between $\tau_\alpha$ and $\tau_\beta$. 
Using reported values of $\psi$ for the different
models (see SI for the details), one obtains the values of $\gamma$ to be
close to $0.51, 0.79, 0.66$ and $0.54$ for 3dKA, 3dHP, 3dR10 and 3dIPL 
models. These numbers are in agreement with the chosen value of 
$\gamma = 0.6$ for three dimensional models. However, the exponent turns 
out to be somewhat different for 3dBMLJ\_82 model ($\gamma \simeq 1.125$). 
We believe that this discrepancy is probably due to uncertainties in 
the values of $\psi, z$ as well as $X$.
For two dimensional models, the numbers are $0.43, 0.58, 0.57$ and $0.54$ 
for 2dmKA, 2dR10, 2dPoly, and 2dKA respectively. Thus the numbers are in 
agreement with the universal number $\gamma = 0.45$ for two dimensions.

If the proposed 
universal relation between $\tau_\alpha$ and $\tau_\beta$ is shown to be 
valid for experimentally relevant glass forming liquids, then one might be
able to understand the vitrification in liquids by probably understanding
the $\beta$-relaxation processes only. This might lead us to identify 
the relevant elementary relaxation process responsible for both $\beta$ 
and $\alpha$ relaxation in glassy systems. On a slightly different note,
it was shown in \cite{MMK2017} that the collapsing dynamics of a polymer 
chain in a supercooled liquid is controlled by both $\alpha$ and $\beta$ 
relaxation processes with possible implications in bio-preservation. It 
is suggested that not only $\alpha$-relaxation but also $\beta$-relaxation 
should be taken into account in order to understand the degradation process 
of biomolecules \cite{CiceroneDouglasBioPhyJ2004, CiceroneDouglasSoftMatter2012}. 
Thus our proposed universal relation between these two relaxation processes 
might help us design the appropriate glassy matrix in future to preserve
bio-macromolecules more efficiently.

To conclude, we have shown that there is a universal relation between 
$\alpha$ and $\beta$ relaxation times of glass forming liquids. The 
proposed relation is different from the one predicted by Coupling Model 
\cite{NGAI98}. The new relation can be rationalized within the framework 
of Random First Order Transition Theory. In two dimensions, due to long 
wavelength density fluctuations, different transport quantities 
show logarithmic system size dependence and disentangling this effect from
the effect emanating from glass transition is often difficult. 
We show how $\tau_\beta$ can be calculated in two dimensions by 
appropriately modifying the correlation functions to remove the effect of long 
wavelength phonon without affecting the cooperative motions at the 
relevant dynamical heterogeneity length scale. We then shown that 
finite size scaling of $\tau_\beta$ is controlled by the dynamic 
heterogeneity length scale as in the three dimensional models. 
Finally, the obtained universal relationship between $\tau_\alpha$ and 
$\tau_\beta$ in both the dimensions suggests that the physics of glass 
transition may be very similar in both two and three dimensions. This
observation is in agreement with recent findings \cite{weeksPNAS, KeimPNAS, 
gilles2d3d}. As $\beta$-relaxation plays in important role below the 
glass transition, in future it will be very interesting and important 
for industrial applications to study possible aging behavior of 
$\tau_\beta$ and its correlation with the aging behavior of $\tau_\alpha$.   

\acknowledgements
We would like to thank Chandan Dasgupta for many useful discussions and 
suggestions.

\end{document}

\title{Possible Universal Relation Between Short time $\beta$-relaxation and 
Long time $\alpha$-relaxation  in Glass-forming Liquids -- Supplementary Information}
\author{Rajsekhar Das\footnote{These authors contributed equally.}}
%\email{rajsekhard@tifrh.res.in}
\author{Indrajit Tah$^{*}$}
%\email{indrajittah@tifrh.res.in}
\author{Smarajit Karmakar}
\email{smarajit@tifrh.res.in}
\affiliation{
Tata Institute of Fundamental Research, 
36/P, Gopanpally Village, Serilingampally Mandal,
Ranga Reddy District, Hyderabad, 500107, India}

\maketitle
We have arranged the supplementary information as follows.
First we will give the details of the models that we have studied and 
discuss about the methods that we have followed in the Sec.~\ref{systemsSI}. 
Next we will discuss how we have calculated $\tau_\beta$ from 
the log derivative of MSD in Sec.~\ref{caltaubetaSI} . In 
Sec.~\ref{crMSDSI} we will discuss about the cage-relative MSD (crMSD) and 
$\tau_\beta$ calculation from crMSD.

\section{Models and Methods \label{systemsSI}}
We have studied the following model glass-forming liquids in both two 
and three dimensions.
 
{\bf 3dKA} -- The well-known Kob-Andersen model \cite{95KA} is a $80:20$ 
binary mixture of two type of  particles where the interacting potential is,
\begin{equation}
\label{LJ}
  V_{\alpha\beta}(r)=4\epsilon_{\alpha\beta}\left[\left(\frac{\sigma_{\alpha\beta}}{r}\right)^{12}-
    \left(\frac{\sigma_{\alpha\beta}}{r}\right)^{6}\right].
\end{equation}
Here the parameters are, $\alpha,\beta \in \{A,B\}$ and $\epsilon_{AA}=1.0$, $\epsilon_{AB}=1.5$,
$\epsilon_{BB}=0.5$, $\sigma_{AA}=1.0$, $\sigma_{AB}=0.80$,
$\sigma_{BB}=0.88$, number density $\rho = 1.20$. We have chosen the cut off 
of the interaction potential as $2.50\sigma_{\alpha\beta}$. We have used a 
quadratic polynomial in such a way that the potential and its first two 
derivatives are continuous at the cut off. Our studied temperature range is 
$T\in \{0.45,3.0\}$. We have done simulations in three dimensions with 
$N = 108000$ number of particles.

{\bf 2dKA} -- This is same model as 3dKA but in two spatial 
dimensions. The temperature range is $T \in \{0.930,2.000\}$. The 
system size is $N = 10000$. This model is known to show prominent medium 
range crystalline order (mrco) at lower temperatures \cite{ISSCS17}.

{\bf 2dmKA} -- This is a slightly modified version of 2dKA model. 
The parameters are same as 2dKA but bigger to smaller particle number 
ratio is $65:35$. Temperature range is $T\in \{0.45,2.0\}$. This model 
is shown to have less tendency towards local crystalline order at lower 
temperatures. 

{\bf 3dR10} -- This is a $50:50$ binary mixture \cite{12KLP} in three 
dimensions. The particles interacts via the potential,
\begin{equation}
\label{R10}
  V_{\alpha\beta}(r) = \epsilon_{\alpha\beta}\left(\frac{\sigma_{\alpha\beta}}{r}\right)^{10}.
\end{equation}
The parameters are, $\epsilon_{\alpha\beta} = 1.0$, $\sigma_{AA} = 1.0$,
$\sigma_{AB} = 1.22$ and $\sigma_{BB} = 1.40$. The potential is cut off 
at $1.38\sigma_{\alpha\beta}$ where we use a  similar quadratic 
polynomial so that the potential and its first two derivatives are 
continuous at the cutoff radius. The number density is $0.81$. Studied 
temperature range is $T \in \{0.52,2.0\}$.

{\bf 2dR10} -- This model is same as 3dR10 model but in two dimensions. 
The number density is $0.85$ and the studied temperature range is 
$T \in \{0.520,2.000\}$. This model is also shown to have almost no 
indication of local ordering at lower temperature. 

{\bf 3dIPL} -- This three dimensional binary mixture \cite{PSD2010} 
is almost similar 
to 3dKA model in terms of parameter of the model but the interaction 
potential is purely repulsive  as given below,
\begin{equation}
V_{\alpha\beta}(r) = 1.945\epsilon_{\alpha\beta}\left(\frac{\sigma_{\alpha\beta}}{r}\right)^{15.48}.
\end{equation}
All the parameters are same as that of 3dKA model. The interaction range 
is larger than the other models with purely repulsive inter-particles 
interaction, like 3dR10 model. The temperature range studied is 
$T \in \{0.45, 1.00\}$. 

{\bf 3dHP} -- This model is a three dimensional $50:50$ binary mixture of 
harmonic spheres \cite{OLLN2002,blockprl} where the diameter ratio of the 
two type of particles is $1.4$.	
This is a model which has been studied extensively in the context of jamming 
in granular medium. In this model, particles interacts via,
\begin{equation}
V_{\alpha\beta}(r) = \epsilon\left[1- \left(\frac{r}{\sigma_{\alpha\beta}}
\right)\right]^2
\end{equation}
if $r < \sigma_{\alpha\beta}$ and $0$ otherwise. 
$\sigma_{\alpha\beta}= (\sigma_\alpha + \sigma_\beta)/2$. Number 
density $\rho = 0.82$ and $\epsilon = 1.0$. The temperature range studied is 
$T \in [0.0045, 0.0090]$. 

{\bf 3dBMLJ\_82} -- This model system consists of equimolar additive mixtures
\cite{cos2007}. 
The system consists of N = 108000 Particles interact via
the Lennard-Jones potential. Particles of species 2 have a smaller diameter than those of 
species 1 $(\sigma_{22} < \sigma_{11})$. The masses
of the two species are equal $m_1 = 1.0$ and $m_2 = 1.0$  and the size ratio
 $\lambda$ = 0.82, keeping $\sigma_{11} = 1.0$. We studied the system in the temperature range $T \in \{0.670,2.000\}$

{\bf 2dPoly} -- This is a poly-disperse mixture of particles where the 
diameter $\sigma_i$ of the particle $i$ is chosen from a Gaussian 
distribution. The polydispersity parameter is then defined as
\begin{equation}
\Delta = \frac{\sqrt{\langle\delta\sigma^2\rangle}}{<\sigma>},
\end{equation}
where $\delta\sigma = \sigma - <\sigma>$. We have chosen $\Delta  = 11\%.$ 
The interaction potential between a pair of particles is defined as,
\begin{equation}
 V_{ij}(r) = 4\epsilon_{ij}\left[\left(\frac{\sigma_{ij}}{r}\right)^{12} -\left(\frac{\sigma_{ij}}{r}\right)^{6} + \frac{1}{4}\right]
\end{equation}
if $r <2^{1/6}\sigma_{ij}$, else $0$. Here $\sigma_{ij} = (\sigma_i +\sigma_j)/2$.
The temperature range studied is $T \in \{0.450,0.900\}$. We simulated the 
system at packing fraction $\eta = 0.76$.

We have performed NVT molecular dynamic simulations for all the model 
system studied. We have studied different system sizes. For two 
dimensional systems we have gone up to $N = 10000$. For three dimensional 
systems we have gone up to $N = 108000$. For simulation we have made sure 
that the systems are equilibrated before storing data by equilibrating 
the systems for at least $100\tau_\alpha$. For better statistical average 
we have performed $32$ statistically independent simulations for each 
temperature.

\section{Correlation Functions}
\label{corrSI}
\noindent{\bf Overlap Correlation Function :}
The point density correlation function or the overlap correlation function
is defined as 
\begin{equation}
Q(t) = \frac{1}{N}\left<\sum_{i=1}^{N}w(|\vec{r}_t(t) - \vec{r}_i(0)|)\right>.
\end{equation} 
The window function $w(x) = 1.0$ if $x \leq 0.3$ and $0$ otherwise. 
This window function is used to remove possible decorrelation that 
might happen due to vibrational motion inside the cage formed around 
a particle by their neighbors.  $\langle \ldots\rangle$ represents 
ensemble average. The $\alpha$ relaxation time is defined as 
$Q(\tau_\alpha) = 1/e$.

\vskip +0.5cm
\noindent{\bf Four-point Susceptibility :}
The four-point susceptibility is defined as the fluctuations of the
overlap correlation function as
\begin{equation}
\chi_4(t) = N\left[\left< Q^2(t)\right> - \left<Q(t)\right>^2\right],
\end{equation}
and the peak of $\chi_4(t)$ appears at timescale close to 
$\alpha$-relaxation time. The peak, $\chi_4^P$, is defined as
\begin{equation}
\chi_4^P \equiv \chi_4(t = \tau_\alpha).
\end{equation}

\vskip +0.5cm
\noindent{\bf Cage relative Mean Squared Displacement (crMDS):}
Mean squared displacement (MSD) is defined as  
\begin{equation}
\langle |\Delta r(t)|^2 \rangle = \left\langle \frac{1}{N}\sum_{i=1}^N 
|\vec{r}_i(t) - \vec{r}_i(0)|^2\right\rangle.
\end{equation}
In two dimensions, MSD will be affected by the long wavelength density 
fluctuations and to remove this effect, a relative MSD with respect 
to cage is calculated. This is termed as cage relative MSD (crMSD). 
It is defined as follows. First we define the cage related 
displacement of particle $i$ as
\begin{equation}
{\Delta}r_{i,CR}(t) = {\Delta}r_i(t) -\frac{1}{N_{nn}}{\sum_{j\in n.n} {\Delta}r_j(t)},
\end{equation}
where $N_{nn}$ are the number of nearest neighbor of $i^{th}$ particle, 
and ${\Delta}r_j(t) = r_j(t) - r_j(0).$ Particles are defined as the 
neighbors of $i^{th}$ particle if they satisfy 
$r_{ij}(t) = |r_j(t) - r_j(0)|< A\sigma_{\alpha\beta}$. 
$A$ is the cutoff value used to define the neighbors. Often one uses 
the value at which first minimum of the radial distribution function 
$g(r)$ appears. Now the crMSD is defined as,
\begin{equation}
\left\langle |{\Delta}r_{CR}(t)|^2 \right\rangle = \left\langle 
\frac{1}{N} \sum_{i = 1}^{N} |{\Delta}r_{i,CR}(t)|^2 \right\rangle
\end{equation} 
We have found that the value of $\tau_\beta$ and particularly the system 
size dependence of it depends crucially on the choice of the cutoff. Below 
we have done a systematic analysis to understand this and to choose an 
appropriate cutoff for the subsequent analysis. 

\section{Calculation of $\tau_\beta$}
\label{caltaubetaSI}
\begin{figure}[!h]
\begin{center}
%\centering
\includegraphics[scale=0.35]{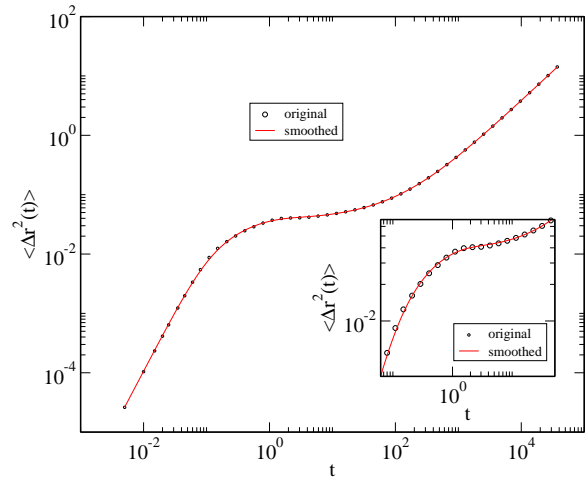}
\caption{Showing original and smoothed MSD data to obtain derivative 
for 2dR10 model for $T = 0.520$. Inset shows the zoomed version of 
the plot to show how small fluctuation in data is avoided. }
\label{smooth}
\end{center}
\end{figure}

\begin{figure}[!h]
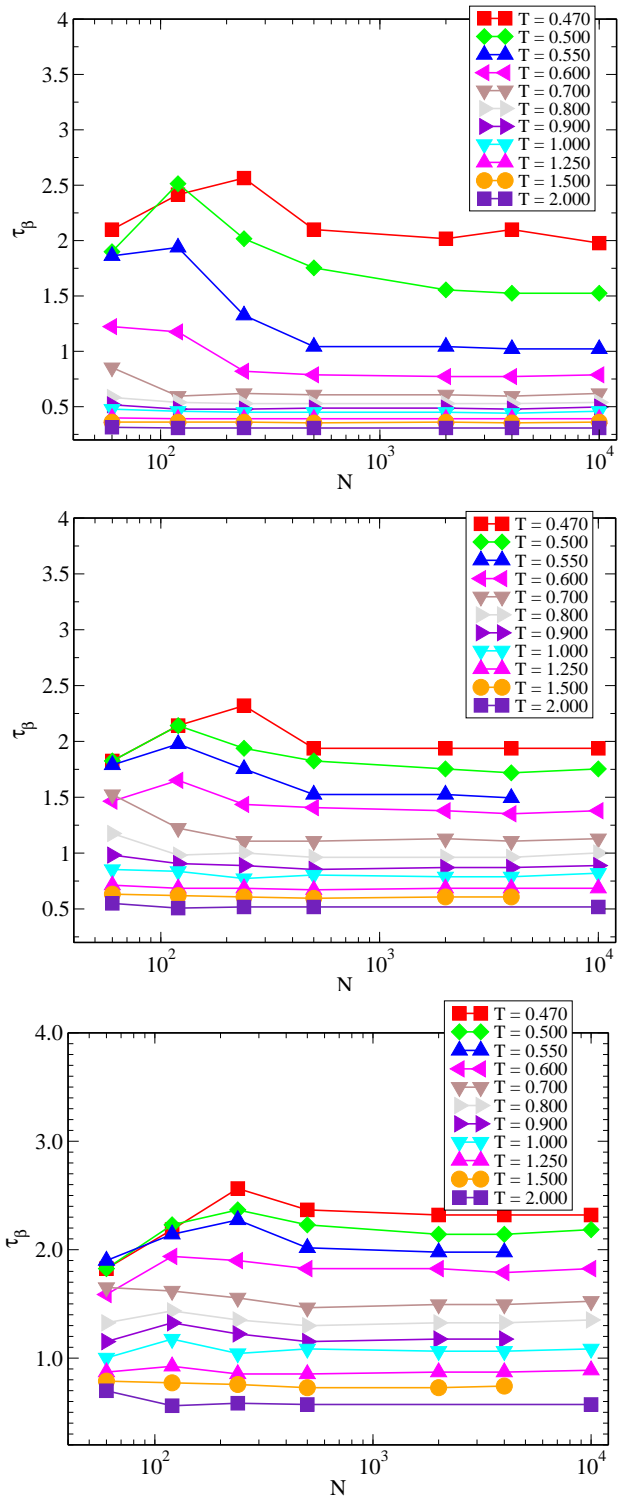

\begin{center}
\centering
\hskip -0.4cm
\includegraphics[width=0.45\textwidth]{cutoff0neSystemsize.eps}
\vskip 0.2cm
\hskip -0.4cm
\includegraphics[width=0.45\textwidth]{cutoffthreesystemsize.eps}
\vskip 0.02cm
\hskip -0.6cm
\includegraphics[width=0.45\textwidth]{cutofffivesystemsize.eps}
\caption{System size dependence of $\tau_\beta$ evaluated by taking logarithmic 
derivative of crMSD (see text for details) for 2dmKA model. Top panel: First 
minimum in pair correlation function $g(r)$ is chosen to define the neighbors 
while calculating crMSD. Middle panel: Same analysis with third minimum as 
cutoff. Bottom panel: using fifth minimum as cutoff distance.}
\label{sysDepSI}
\end{center}
\end{figure}
\hskip 0.3cm
As discussed in the main text we have calculated the $\beta$-time scale 
$\tau_\beta$ from the minimum of the dlog$<|\Delta r(t)^2|>$/dlog(t) vs 
$log(t)$ plot. To calculate numerical derivative as we need very closely 
spaced data, we have used a splined data. We used cubic spline to get 
the splined data. This splined data was again smoothen before calculating 
the derivative as a minute fluctuation in the data will cause a huge 
fluctuation in derivative, which will make it difficult to find the 
minimum unambiguously. In fig.~\ref{smooth} we have shown the original 
and smoothed MSD for 2dR10  model for $T=0.520$. The inset shows the 
plateau region zoomed to show the quality of the spline interpolation. 
Similar method is also used for extracting the value of $\tau_\beta$
for two dimensional models from crMSD. 

\section{Cutoff dependence of crMSD}
\label{crMSDSI}
We have discussed in the main text that the system size dependence of 
$\tau_\beta$ depends on the choice of the cutoff we while calculating 
crMSD. If we choose 2nd neighbor distance as cutoff instead of 1st 
neighbor distance, then the system size dependence of $\tau_\beta$ 
changes. In Fig.~\ref{sysDepSI} we show the system size dependence of 
$\tau_\beta$ for three different cutoff distances. One can clearly 
see how system size dependence changes on the choice of cutoff for 
2dmKA model. As this becomes somewhat ambiguous, we choose dynamic
heterogeneity length scale as the cutoff. With this choice, the values
of $\tau_\beta$ seem to show behaviour very similar to the results 
obtained for three dimensional models. In the main articles, all 
reported numbers of $\tau_\beta$ for two dimensional models are 
obtained with dynamic heterogeneity length scale as cutoff for 
neighbor calculation.    

\section{Effect of system size on the relation between $\tau_\beta$
and $\tau_\alpha$: }
\begin{figure}[!h]
\begin{center}
%\centering
\includegraphics[scale=0.42]{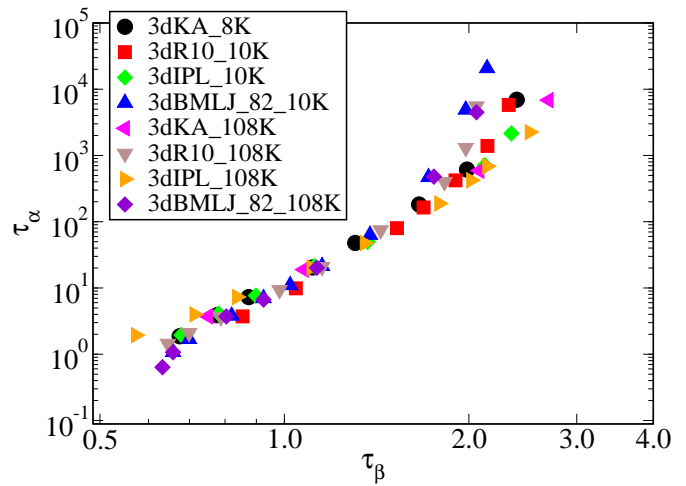}
\caption{Log-log plot of $\tau_\alpha$ vs $\tau_\beta$ for different 
model systems in three dimensions with data obtained for different
system sizes. One can see that deviation from a universal power law 
relation becomes much larger for smaller system sizes.}
\label{sysEffRel}
\end{center}
\end{figure}
In Fig.\ref{sysEffRel}, we have plotted $\tau_\alpha$ as a function 
of $\tau_\beta$ for all the model systems in three dimensions using
data from different system sizes. The deviation observed from a 
possible universal power law relation is much stronger than with 
the data for only larger system sizes. 

%\begin{widetext}
\onecolumngrid
\section{Exponents $\psi$, $X$, $z$ and $\gamma$:}
The details of the exponents ($\psi$, $X$, $z$ and $\gamma$) 
quoted in the main articles are given below:

\parbox{.45\linewidth}{
\begin{center}
\begin{tabular}{ |c|c|c|c|c| }
 \hline 
\multicolumn{5}{|c|}{Three Dimensional Models}\\
 \hline
  & $\psi$ & $z$ & $X$ & $\gamma = \psi/zX$ \\
  \hline 
 \,3dKA \,& \,1.10\, & \,0.80\, & \,2.70\, & \,0.51\, \\
  \hline 
 3dHP & 2.22 & 0.80 & 3.50 & 0.79 \\ 
 \hline
 3dR10 & 1.86 & 0.80 & 3.50 & 0.66\\
 \hline
 3dIPL & 1.21 & 0.80 & 2.80 & 0.54 \\
 \hline
 \,3dBMLJ\_82\, & 0.90 & 0.80 & 1.00 & 1.125\\
 \hline
\end{tabular}
\end{center}}
\hfill
\parbox{.45\linewidth}{
\begin{center}
\begin{tabular}{ |c|c|c|c|c| }
 \hline 
\multicolumn{5}{|c|}{Two Dimensional Models}\\
 \hline
  & $\psi$ & $z$ & $X$ & $\gamma = \psi/zX$ \\
  \hline 
 \,2dKA \,& \,0.68\, & \,1.25\, & \,1.00\, & \,0.54\, \\
  \hline 
 2dmKA & 0.70 & 1.25 & 1.30 & 0.43 \\ 
 \hline
 2dR10 & 0.89 & 1.25 & 1.23 & 0.58\\
 \hline
 2dPoly & 0.71 & 1.25 & 1.00 & 0.57 \\
 \hline
\end{tabular}
\end{center}}
%\end{widetext}

\bibliography{$HOME/Dropbox/docs/jShort.bib,$HOME/Dropbox/docs/mybiblio.bib}